\documentclass{elsart3}

 \usepackage{graphicx}

\usepackage{amssymb}

\begin{document}

\begin{frontmatter}



\title{Magnon softening and damping in the
ferromagnetic manganites due to orbital correlations}


\author[1]{S. Krivenko},
\author[2]{A. Yaresko},
\author[3]{G. Khaliullin} and
\author[4]{H. Fehske\corauthref{cor1}}
\address[1]{Dipartimento di Fisica,
Unit\`{a} di Ricerca INFM di Salerno  Universit\`{a} degli Studi
di Salerno, I-84081 Baronissi (SA), Italy}
\address[2]{Max-Planck Institut f\"ur Chemische Physik fester
Stoffe, D-01187 Dresden, Germany}
\address[3]{ Max-Planck Institut f\"ur Festk\"orperforschung, D-70569
Stuttgart}
\address[4]{Institut f\"ur Physik, Ernst-Moritz-Arndt Universit\"at
Greifswald, D-17487
Greifswald, Germany}

\corauth[cor1]{Corresponding author. Fax: +49 3834 864701; E-mail:
fehske@physik.uni-greifswald.de}

\begin{abstract}
We present a theory for spin excitations in ferromagnetic metallic
manganites and demonstrate that orbital fluctuations have strong
effects on the magnon dynamics in the case these compounds are
close to a transition to an orbital ordered state. In particular
we show that the scattering of the spin excitations by low-lying
orbital modes with cubic symmetry causes both the magnon softening
and damping observed experimentally.

\end{abstract}

\end{frontmatter}

Metallic ferromagnetic manganites belong to the class of double
exchange systems
, in which the motion of doped
charge carries establishes a ferromagnetic interaction between
neighboring Mn
spins.
In some of the colossal magnetoresistive
manganese perovskites
, the
spectrum of magnetic excitations (magnons) has a rather unusual
shape: while showing standard Heisenberg behavior at small
momenta the magnon dispersion exhibits a pronounced softening near
the Brillouin zone boundary. In addition, magnons with short wave
lengths cease to be well defined quasiparticles because width of
their spectral density peaks become comparable with their energy
at the Brillouin zone boundary.~\cite{Dai00,Dai99,Hwang}
Theoretically, Furukawa~\cite{FUR} has explained this observation
by the scattering of the magnons due to the
optical phonons.
Khaliullin and Kilian ~\cite{KhKi00}
attributed the effect to the orbital fluctuations.

The effective
Hamiltonian of scattering of the magnons $b_i$ by the orbital
excitations of the $e_g$ electrons, expressed by $f_{i\alpha}$
-fermions has the form~\cite{KhKi00}:
\begin{eqnarray}\label{hstart}
  H =
- \sum_{\langle i,j \rangle_{\gamma}} \sum_{\alpha \beta} &&
xt^{\alpha \beta}_{\gamma} f_{i\alpha}^{\dagger}f_{j\beta}^{}[1
\nonumber\\ &-& \frac{1}{4S}(b_{i}^{\dagger}b_{i}^{}
+b_{j}^{\dagger}b_{j}^{} -2 b_{i}^{\dagger}b_{j})].
\end{eqnarray}


In the present work, we
introduce  also a direct effective
interaction between the $e_g$ orbitals of the nearest-neighbor
(NN) manganese sites
\begin{equation}
\label{horba} H_{orb}= V\sum_{\langle
ij\rangle_{\gamma}}\tau^{(\gamma)}_{i}\tau^{(\gamma)}_{j}\,,
\end{equation}
where the orbital isospins defined as $\tau^{(z)}=\sigma^z/2$\,,
$\tau^{(x/y)}=-(\sigma^z\pm\sqrt{3}\sigma^x)/4$, and  $\sigma^z$
and $\sigma^x$ are Pauli operators acting in the orbital sector.
In the classical limit, the Hamiltonian~(\ref{horba}) can be
diagonalized in the reciprocal space giving
two normal modes
$\sigma^{(+)}_{\textbf{\textit{q}}}$ and
$\sigma^{(-)}_{\textbf{\textit{q}}}$, with
dispersion relations
\begin{equation}
\label{worb} \omega^{\pm}_{\textbf{\textit{q}}}= \omega_0\Bigl(
1+(\gamma_{\textbf{\textit{q}}} \pm
\sqrt{\eta_{2\textbf{\textit{q}}}^2+\eta_{3\textbf{\textit{q}}}^2})\,\textrm{sign}(V)
\Bigr)\,,
\end{equation}
where $\omega_0=3|V|/8$,
$\eta^{(2)}_{\textbf{\textit{q}}}=\sqrt{3}(c_y-c_x)/6$,
$\eta^{(3)}_{\textbf{\textit{q}}}=(2c_z-c_x-c_y)/6$ and
$\gamma_{\textbf{\textit{p}}}=(c_x+c_y+c_z)/3$.


An effective Hamiltonian describing the interaction between the
magnons and the low-lying collective orbital mode can be derived
by combining the spin-orbiton coupling in Eq.(\ref{hstart}) with
the direct orbital-coupling term~(\ref{horba}), leading to
\begin{equation}
\label{hsct}
H_{s-orb}=\sum_{\textbf{\textit{p}}\textbf{\textit{q}}}
\Bigl(g^+_{\textbf{\textit{p}}\textbf{\textit{q}}}\sigma^{(+)}_{-\textbf{\textit{q}}}
+g^-_{\textbf{\textit{p}}\textbf{\textit{q}}}\sigma^{(-)}_{-\textbf{\textit{q}}}\Bigr)
b^{\dagger}_{\textbf{\textit{p}}}b_{\textbf{\textit{p}} +
\textbf{\textit{q}}}\,.
\end{equation}

The magnon spectrum is given by
$\tilde\omega_{\textbf{\textit{p}}}=\omega_{\textbf{\textit{p}}}+
\textrm{Re}[\Sigma(\omega_{\textbf{\textit{p}}},
\textbf{\textit{p}})]\,$ with
the mean-field dispersion
$\omega_{\textbf{\textit{p}}}=zD(1-\gamma_{\textit{p}})$
(resulting from $\langle f_{i\alpha}^{\dagger}
f_{j\beta}^{}\rangle$ \cite{KhKi00}).
$\Sigma(\omega,\textbf{\textit{p}})$
denotes the magnon self-energy, where
$\tilde\Gamma_{\textbf{\textit{p}}}=-
~\textrm{Im}[\Sigma(\omega_{\textbf{\textit{p}}},\textbf{\textit{p}})]$
is the magnon damping.
The magnon self-energy is the sum of two contributions,
stemming from orbiton particle-hole excitations and
from the orbital collective mode.

\begin{figure}
 \centering \includegraphics[height=4.4cm, width=5cm]{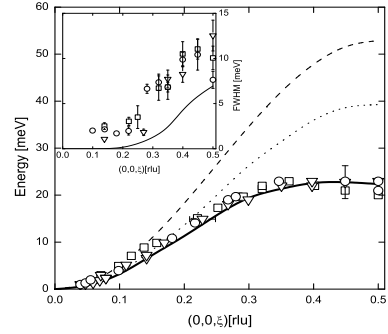}
 \caption{Magnon dispersion along the $(0,0,\xi)$-direction.
The dashed line corresponds to the function
$\omega_{\textbf{\textit{p}}}$. The solid (dotted) line gives
$\tilde\omega_{\textbf{\textit{p}}}$ with (without) contribution
from the orbital collective mode.
Inset: Magnon width $2
\tilde\Gamma_{\textbf{\textit{p}}}$.
Here squares, triangles and circles give experimental
data~\protect\cite{Dai00} for Pr$_{0.63}$Sr$_{0.37}$MnO$_3$, $\rm
Nd_{0.7}Sr_{0.3}MnO_3$, and La$_{0.7}$Ca$_{0.3}$MnO$_3$,
respectively. }
 \label{fig2}
\end{figure}

\begin{figure}
 \centering \includegraphics[height=4.4cm, width=5cm]{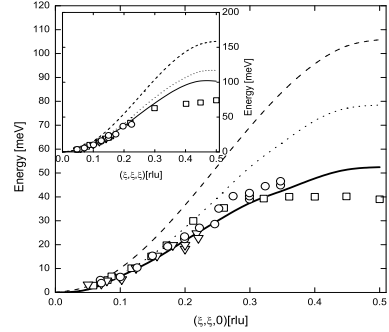}
 \caption{Magnon dispersion along the $(\xi,\xi,0)$- [$(\xi,\xi,\xi)$-]
direction [inset]. Line styles are the same as in
Fig.~\protect\ref{fig2}. Experimental data in main figure
is taken from~\protect\cite{Dai00}.
The Pr- (Nd-, La-) data shown in the inset is taken
from~\protect\cite{Hwang}~(\protect\cite{Dai99}).} \label{fig3}
\end{figure}

Calculated magnon
dispersions are displayed in Figs.~\ref{fig2} and~\ref{fig3}. The
inset of Fig.~\ref{fig2} shows 
the momentum dependence of magnon width.
Scattering of the magnons due to the excitations of the
orbiton particle-hole pairs does not change the Heisenberg type of
the spin-wave dispersion (it results only in a partial decrease of
their bandwidth [dotted lines on Figs.~\ref{fig2} and
~\ref{fig3}]).
On contrary, the development of the soft
orbital modes affects the spin dynamics substantially. Their
effect is particularly pronounced  for the spin excitations with
momentum $\bf{p}$  along the direction $(0,0,\xi)$,
where the magnon dispersion  flattens as one approaches the
Brillouin zone boundary. At the same time, the magnon width
increases abruptly (see inset in Fig.~\ref{fig2}).
The effect of the collective modes is
small for the direction $(\xi,\xi,\xi)$ (see Fig.~\ref{fig3}).
This anisotropic behavior
originates from the
peculiar structure of the dispersion in Eq.~(\ref{worb}),
exhibiting soft lines in direction $(0,0,\pi)$ and equivalent ones.

To summarize, 
a strong damping of the short
wave length magnons and a marked deviation of their spectrum from
a canonical Heisenberg form
may originate from the scattering of the magnons by a
collective orbital mode, and therefore can be understood as a
precursor effect of orbital ordering.

SK thanks the Grant RFFI-03-02-17453.


%




\end{document}